\documentstyle[aps,12pt]{revtex}

\begin{document} \begin{center}

{\bf \huge {Intermittency in Inhomogeneous Coupled Map Lattices} } \end{center}
\vskip 1.0in \begin{center}

Ashutosh Sharma$^{1,2}$ and Neelima Gupte$^2$ 

\vskip 0.1in $^1$Department of Physics,

University of Pune,

Pune ---411007, INDIA.  \vskip 0.1in

{$^2$ Department of Physics,

Indian Institute of Technology,

Madras 600036, INDIA }

\end{center} \vskip 0.5in 
\begin{abstract}
We study the phenomenon of intermittency in an inhomogeneous
lattice of coupled maps where the inhomogeneity appears in the form of
different values of the 
map parameter at adjacent sites. 
This system exhibits  spatio-temporal intermittency over substantial regions of
parameter space. We also observed the unusual feature of purely spatial
intermittency accompanied by temporal periodicity in some regions of the
parameter space.  
The
intermittency appears as a result of bifurcations of co-dimension two in
such systems. We identify the types of bifurcations which lead to
spatio-temporal or purely spatial intermittency and identify the mechanism of intermittency
for
one such case, that arising from the bifurcation of the synchronised
fixed point solution, by examining the spatial return map. The
intermittency near the bifurcation points is associated with power-law
distributions for the laminar lengths. The scaling laws for 
the laminar 
length distributions are obtained. Three distinct types of intermittency
characterised by power-laws with three exponents which fall in three distinct ranges can be seen.  
The structure associated with the tangent-period-doubling bifurcation
undergoes further bifurcation to a spatio-temporally intermittent regime
with an associated power-law exponent. Three of the exponents seen
in this model show very good agreement with  
those observed in fluid experiments involving quasi-one dimensional geometries.

\end{abstract}
 
\newpage

The existence of spatio-temporal as well as temporal intermittency is 
an interesting phenomenon which is found in a
wide range of dynamical systems of physical interest such as
oscillators\cite{osc},
chemical reactions \cite{chem}, pattern formation of various kinds
\cite{patt},
turbulence \cite{turb}, fluid flows \cite{Colovas} and signals of all kinds. The phenonomenon of temporal
intermittency has been
studied extensively and is relatively well understood \cite{berge}.
Pomeau and Manneville \cite{pomeau} first established the  route to
chaos via temporal intermittency and studied its behaviour. The nature
of intermittency in spatially extended dynamical systems, however has
not
been understood very  well. The presence of spatial as well as temporal
intermittency
has implications for understanding the physics of pattern formation and
for understanding
the ubiquitous presence of structures in chaotic systems.
 
Coupled map lattices are particularly simple paradigms for
systems with extended spatial dimension which show a wide range of
interesting behaviour
\cite{kaneko}. Spatio-temporal intermittency, which has been defined as a
`fluctuating mixture of regular and turbulent domains \cite{chate}',  has been observed in such
systems \cite{kaneko}\cite{chate}\cite{gang}. However, the mechanism by
which such intermittency
occurs has not been very clear.
We study the phenomenon of spatio-temporal intermittency in an inhomogeneous
lattice of coupled maps where the inhomogeneity appears in the form of 
different values of map parameter at different  sites. 
Such lattices have been considered
in the case of pinning studies\cite{pinn} and in the context of control of
spatio-temporal chaos \cite{gang}.  The coupling between maps is diffusive and nearest
neighbour.  We consider the simplest case, that is, a situation where adjacent
lattice sites have different values of the parameter and alternate sites
have the same value of the parameter. 

Our model is defined by the evolution equations \begin{equation}
     x_{i}^{t+1}=(1-\epsilon)f(x_{i}^{t}, \mu)+{\epsilon\over 2}{(f(x_{i-1}^{t},
     \mu^{\prime}) +f(x_{i+1}^{t}, \mu^\prime))} \label{cml} \end{equation}

where $f(x)= \mu x(1-x)$ is the logistic map and $\mu$, $ \mu^\prime \epsilon$
 $[0,4]$, $ x_{i}^{t}$ is the value of the variable $x$ at the lattice site
$i$ at time $t$, and $ 0 \leq x \leq 1$. 
We set $\mu^\prime=\mu-\gamma$ and  use periodic
                   boundary conditions where $2N$, the number of lattice sites is even.
The synchronised fixed points of the system are given by $x=0$ and
$x={{\mu-\gamma\epsilon-1}\over{\mu-\gamma\epsilon}}$

Expanding the evolution equations about the synchronised fixed point, the linear stability matrix has the form 

\begin{equation} J={\left[{\begin{array}{ccccccc}(1-\epsilon)f^{\prime}_{\mu}(x)
      &{\epsilon\over 2}f^{\prime}_{\mu^{\prime}}(x) &0&0&.  .  .
      &0&{\epsilon\over 2}f^{\prime}_{\mu^{\prime}}(x)\\ {\epsilon\over
      2}f^{\prime}_{\mu}(x)&(1-\epsilon)f^{\prime}_{\mu^{\prime}}(x)&{\epsilon\over
      2} f^{\prime}_{\mu}(x)&0&.  .  .  &0&0 \\ 0&{\epsilon\over
      2}f^{\prime}_{\mu^{\prime}}(x)&(1-\epsilon)f^{\prime}_{\mu}(x)
      &{\epsilon\over 2}f^{\prime}_{\mu^{\prime}}(x)&.  .  .  &0&0\\
      0&0&{\epsilon\over
      2}f^{\prime}_{\mu^{\prime}}(x)&(1-\epsilon)f^{\prime}_{\mu}(x) &.  .  .
      &0&0\\

.  .  .  &.  .  .  &.  .  .  &.  .  .  &.  .  .  .  &.  .  .  &.  .  .  .  \\ .
                               .  .  &.  .  .  &.  .  .  &.  .  .  &.  .  .  .
                               &.  .  .  &.  .  .  .  \\

0&0&.  .  .  &.  .  .  &.  .  &(1-\epsilon)f^{\prime}_{\mu}(x)&{\epsilon\over
2}f^{\prime}_ {\mu^\prime}(x)\\ {\epsilon\over 2}f^{\prime}_{\mu}(x)&0&.  .  .
&.  .  .  &.  .  &{\epsilon\over 2}f^{\prime}_ {\mu}(x)&
(1-\epsilon)f^{\prime}_{\mu^{\prime}}(x)\end{array}}\right]} \label{lst}
\end{equation}

This matrix is of the  block circulant form. Thus it can be put in a block diagonal form 
by a similarity transformation \cite{Davis},\cite{Gade} 
of blocks  $M(\theta)$ where $M(\theta)$ are $2\times2$ matrices of the form

\begin{equation} M(\theta)={\left[{\begin{array}{cc}
(1-\epsilon)f_{\mu}^{\prime}(x)&{\epsilon\over
2}(1+e^{i\theta})f_{\mu^{\prime}}^{\prime}(x)\\ {\epsilon\over
2}(1+e^{-i\theta})f_{\mu}^{\prime}(x)&(1-\epsilon)f_{\mu^{\prime}}^
{\prime}(x)\end{array}}\right]} \\ \label{2} \end{equation}

where, 
$\theta={{2\pi(l-1)}\over N}$, $l=1, \ldots N$.

The eigenvalues of this matrix as a function of $\theta$ are given by
\begin{eqnarray}
\lambda={{(1-\epsilon)(f_{\mu}^{\prime}+f_{\mu^\prime}^{\prime})\pm{((1-\epsilon)
(f_{\mu}^{\prime}+f_{\mu^\prime}^{\prime})^{2}-2(1-\epsilon)\epsilon(f_{\mu}^{\prime}f_
{\mu^\prime}^{\prime})^{2}(1+cos(\theta)))}^{1/2}}\over
2}
\label{eigen}
\end{eqnarray}

It is clear from Eq. \ref{eigen} that the eigenvalues of the matrices $M(\theta)$ are bounded between
the largest eigen-value of the matrix $M(0)$ and the smallest
eigen-value of the matrix $M(\pi)$. 

      For the $x=0$ fixed point, the eigenvalue equations are \begin{eqnarray}
\lambda^{2}-\lambda(1-\epsilon)(2\mu-\gamma)+\mu(1-2\epsilon)(\mu-\gamma)=0 \hskip 0.2infor
M(0) \nonumber \\
\lambda^{2}-\lambda(1-\epsilon)(2\mu-\gamma)+\mu{(1-\epsilon})^{2}(\mu-\gamma)=0
\hskip 0.2infor M(\pi) \end{eqnarray}

For the fixed point, $x={{\mu-\gamma\epsilon-1}\over{\mu-\gamma\epsilon}}$ the eigenvalue
equations are \begin{eqnarray}
\lambda^{2}-\lambda(1-\epsilon)(2\mu-\gamma)\xi+\mu(1-2\epsilon)(\mu-\gamma)\xi^{2}=0
\hskip 0.2infor M(0) \nonumber \\
\lambda^{2}-\lambda(1-\epsilon)(2\mu-\gamma)\xi+\mu{(1-\epsilon})^{2}\xi^{2}(\mu-\gamma)=0
\hskip 0.2infor M(\pi) \end{eqnarray}
where  $\xi={{\gamma\epsilon-\mu-2}\over{\mu-\gamma\epsilon}}
$.

We fix the value of $\mu$ at $4.0$. The system now has two parameters
$\epsilon$ and $\gamma$. It is clear from Eq. \ref{eigen} that the
eigen-values of the linear stability matrix are functions of $\epsilon
$ and $\gamma$ . 

It is well-known \cite{Guckenheimer} that 
bifurcations  occur and  the nature of the stable solution  changes
at the values of the parameters where the modulus of the eigen-values
of the linear stability matrix 
crosses the unit circle. In the co-dimension one case, the tangent bifurcation can be seen where the
largest real eigen-value crosses $+1$, the period-doubling bifurcation can be
seen where the smallest real eigen-value crosses $-1$, and a Hopf bifurcation
can be seen where the complex eigen-values cross the unit circle. 
For the present system, the conditions for the eigenvalues to cross the unit circle define a set
of 
curves in the two-parameter  $\epsilon-\gamma$ space as listed in Table I.
These curves intersect in several places where the equations are 
simultaneously satisfied. At these intersections two eigen-directions become
unstable, resulting in bifurcations of co-dimension two.  A  rich
variety of 
spatio-temporal behaviour can be seen in the neighbourhood of such points. 
We concentrate on  regions which exhibit spatio-temporal intermittency
where a fluctuating mixture of regular and irregular domains can be
seen. 
Such solutions are possible over a large region of parameter space.
Table I lists the
possible bifurcations associated with intermittent behaviour and the different types of intermittent behaviour observed near such points.
It is interesting to note that in addition to the phenomenon 
of spatio-temporal intermittency, the inhomogeneous lattice has regions in
parameter space where pure spatial intermittency accompanied by
temporally periodic behaviour can be seen i.e. the temporal behaviour is
periodic, but spatially laminar and turbulent regions co-exist at any
point in time (See Fig. 1).
The temporally periodic spatial structure seen shows the unusual occurence of long range spatial correlations.
This phenomenon can be seen in the vicinity of the 
tangent-period doubling bifurcation 
 whereas all the other bifurcations result in
spatio-temporal intermittency. 
The bifurcations in
case of the Double Hopf and period-doubling are inverse bifurcations as can be confirmed by
a normal form analysis.The details of this calculation
 can be found elsewhere \cite{SG}.

Table I shows that spatio-temporal as well as purely spatial intermittency can arise in the
neighbourhood of co-dimension 2 bifurcations.  We examine one such case, that
of the tangent-period doubling bifurcation where
the interesting case of purely spatial
intermittency can be seen.  Let $\chi_{0}(\gamma, \epsilon,
\lambda)$ be the characteristic polynomial for $M(0)$ and $\chi_{\pi}(\gamma,
\epsilon, \lambda)$ be the characteristic polynomial for $M(\pi)$.  Then the
conditions to be simultaneously satisfied for the Tangent-Period doubling
bifurcation are 

\begin{eqnarray}
1-(1-\epsilon)(2\mu-\gamma)\xi+\mu(1-2\epsilon)(\mu-\gamma)\xi^{2}=0 \nonumber
\\ 1+(1-\epsilon)(2\mu-\gamma)\xi+\mu(1-\epsilon)^{2}(\mu-\gamma)\xi^{2}=0
\end{eqnarray}

where $\xi={{\gamma\epsilon-\mu-2}\over{\mu-\gamma\epsilon}} $ as before.  These
                              conditions are simultaneously satisfied when
                              $\gamma=1.  18, \epsilon=0.  63$.

We show the spatial intermittency present in the vicinity of this point in Fig
1(a). The size of lattice chosen was $1000$ and it was iterated for $10,000$
iterations after discarding $20,000$ transients.  The ordinate is the value of x
at the $i$th site and abscissa is the site label.  The lattice is plotted at a fixed time.
In Fig.1(b) we plot the time evolution of the lattice which shows stable
period two oscillations.  Thus the intermittency is purely spatial in nature and
temporally we have stable periodic behaviour.
The temporally periodic spatial structure seen shows the unusual occurrence of long range
spatial correl
ations. This can be seen in the distribution of laminar lengths which we will discuss shortly. 

Another tool that can give some insight into the intermittency for this
Tangent-Period-doubling case  is the spatial
return map plotted in Figure 3. Before the bifurcation, the return map shows only a single point
corresponding to the synchronised fixed point solution. After the bifurcation a
loop is seen. This is because intermediate values between the two period doubled
points are also possible now. This loop acts as a relaminarisation mechanism
and gives rise to intermittency. 
The region in which the intermittency occurs in inhomogeneous lattices is
considerably bigger than that seen for homogeneous lattices.  This is because
 the lattice is no
longer symmetric under the "flip".  Thus the periodic solutions which were
stable for homogeneous lattices bifurcate and give rise to intermittency.

Each type of bifurcation gives rise to intermittency of a
distinct type as typified by the distribution of laminar lengths.
The distribution of laminar lengths shows power law behaviour, and the tangent, the period-doubling and the Hopf bifurcation are associated
with their own distinct power laws. The distributions for the tangent-Hopf and tangent-period-doubling bifurcations scale with the power-laws associated
with the Hopf and period-doubling bifurcations respectively. These power-laws
are seen in the vicinity of the co-dimension two points listed in Table I and also in  the neighbourhood of the lines which correspond to the tangent bifurcations.
The question of whether
spatio-temporal intermittency persists beyond these regions  and whether the distribution of
laminar lengths in the spatio-temporally intermittent regime crosses over from power-law behaviour to exponential behaviour 
is presently under investigation.

To distinguish between various kinds of intermittency, we calculate the
distribution of laminar lengths.  The length of the laminar bursts, i.e. the
number of consecutive sites which follow periodic behaviour before being
interrupted by chaotic bursts is calculated.  The distribution for this length
shows a power law behaviour with $P(l)\approx l^{\zeta}$.  This shows the
presence of long-range spatial correlations in the lattice.  The power law
exponent characterises the type of intermittency.  The distribution of laminar
lengths for three kinds of intermittency is shown in Fig.2. The exponent $\zeta_{1}$ is associated
with the tangent bifurcation,the
exponent  $\zeta_{2}$  with 
the  Tangent-period-doubling  bifurcation and period-doubling
(sub-critical) and the exponent $\zeta_{3}$ with
tangent-Hopf
 and the Double Hopf bifurcations.
Such power-laws have been observed in various experimental studies of spatio-temporal intermittency \cite{Colovas}. The values of the power-laws are listed 
in Table I. It is interesting to note that the values seen by us are in reasonable
agreement with the power-laws seen in several experiments which involve quasi-one dimensional
geometries\cite{Ciliberto},
\cite{Michal}, \cite{Mutabazi}. The exponent
$\zeta_1$ which takes values between $[1.9-2.2]$
for our system, is in good agreement with the laminar
exponent seen  in the case of Rayleigh-Benard convection in an
annulus \cite{Ciliberto}, whereas the exponent
$\zeta_3$ which takes the values in the range $[0.61-0.72]
$ agrees well with the laminar exponent
seen in the case of the roll coating system
\cite{Michal}. The exponent
$\zeta_2$ which takes values in the range  $[1.3-1.35]$ and which is seen in 
the case of the periodic structure with long range spatial correlations 
is a lower bound on the laminar exponents seen in the case of
the Rayleigh-Benard convection seen in a channel and in
the Taylor Dean system where spatio-temporal intermittency is seen \cite{Colovas1}, \cite{Mutabazi}.
Our numerical studies show that this is due to the fact that the temporally periodic structure with long-range spatial correlations undergoes a further bifurcation to  
spatio-temporal intermittency in the neighbourhood of $\epsilon=0.368, \gamma=0.56$ for bifurcations from the synchronised fixed point $x=0.0$. The distribution of laminar lengths in this region 
shows power-law behaviour with an exponent $\zeta_G \approx 1.63$ which is in very good agreement
with the laminar exponents observed for convection in a channel and for the Taylor-Dean system. This bifurcation 
does not appear to be local in nature. Details of this bifurcation are being explored further.

All the results above are obtained for bifurcations from the
synchronised fixed points for a lattice where the map parameter takes
two distinct values one at each alternate site. These results can be
easily generalised to higher spatial and temporal periods and to
lattices where the inhomogeneity has different periodicities. Bifurcations
from higher spatial temporal periods can also give rise to spatio-temporal
or spatial/temporal intermittency in many additional regions of parameter
space. The question of whether
the spatio-temporal intermittency with or without the accompanying
power-laws for the distribution of laminar lengths persists in other regions of parameter space is being investigated further\cite{SG}.

Thus we have shown that both spatial and
spatio-temporal intermittency can arise in a
inhomogeneous coupled map lattice. 
The phenomenon of intermittency is
more widespread in inhomogeneous lattices than in the case of ordered lattices. The
presence of pure spatial intermittency accompanied by temporally
periodic behaviour is an interesting feature which
arises in the case of the inhomogeneous system.
The intermittency arises as a result of bifurcations of co-dimension 2.
Such bifurcations are also of interest in the case of other spatially
extended systems
 \cite{Turing-Hopf}.
The distributions of laminar lengths exhibit three
distinct kinds of power laws, each associated with a distinct kind of
bifurcation. The structure associated with the tangent-period-doubling
bifurcation undergoes a further bifurcation associated with an
additional power-law.
 The values obtained for these
power-laws are in reasonable and intriguing agreement with
those observed in a variety of experiments
involving quasi-one dimensional geometries.
We hence hope our analysis will be
useful for the understanding of intermittent phenomena arising in other
spatially extended systems, and in discussions of their genericity
and universality. 

The authors  thank the Institute of Mathematical Sciences, Chennai , for
their computational facilities and  Prof. R. Ramaswamy for
a  very useful discussion. AS thanks
Dr. Prashant Gade for invaluable discussions.  This work was carried
out under grants from the University Grants Commission, India, and the Department of Science and
Technology, India (Grant no. SP/S2/E-03/96).

\newpage 
\begin{center}
{\bf TABLE 1.1}\\
\vspace{0.15in}
{ Fixed Point x= 0 } \\ 
\end{center}
\vskip 0.1in
\begin{tabular}{|l|l|l|l|l|}
 \hline Type & Parameter region
&Eigenvalue &$P(l)\approx
l^{\zeta}$ &Nature \\ \hline Tangent
Bifurcation&$\epsilon={(5\gamma-21)\over(9\gamma-40)},
{(3\gamma-9)\over(7\gamma-24)}$&+1& $\zeta_{1}$&ST \\ \hline Tangent-PD
Bifurcation&$\epsilon=0.  39, \gamma=0.  56$&+1, -1& $\zeta_{2}$&Spatial \\ \hline
Tangent-Hopf Bifurcation&$\epsilon=0.  245, \gamma=2.  87$&$a\pm ib, +1$&$\zeta_
{3}$
&ST \\ \hline\end{tabular}
\vskip 0.5in
\begin{center}
{\bf TABLE 1.2}\\
\vspace{0.15in}
{ Fixed
Point $x={{\mu-\gamma\epsilon-1}\over{\mu-\gamma\epsilon}}$} \\
\end{center}
\vskip 0.1in
 \begin{tabular}{|l|l|l|l|l|} \hline Type & Parameter region
&Eigenvalue &$P(l)\approx
l^{\zeta}$ &Nature \\  \hline  Tangent
Bifurcation&$\epsilon={{5\gamma+8\pm
(5\gamma-192\gamma+64)^{1\over2}}\over{8\gamma}}$&+1& $\zeta_{1}$&ST \\ \hline P
D
Bifurcation&$\epsilon={{\gamma^{2}-3\gamma-8\pm
{({(\gamma^{2}-3\gamma-8)^{2}-8\gamma(\gamma-4) (\gamma-6))}
^{1\over2}}\over{(2\gamma(\gamma-4))}}}$&+1& $\zeta_{2}$&ST \\ \hline  Tangent -
PD
Bifurcation&$\epsilon=0.  63, \gamma=1.  180$&-1, +1& $\zeta_{2}$&Spatial \\ \hline
Tangent-Hopf Bifurcation&$\epsilon=0.  06, \gamma=2.  9801$&$+1, a\pm ib$&
$\zeta_{3}$&ST\\ \hline Double-Hopf Bifurcation&$\epsilon=0.  5, \gamma=2$&$a\pm
ib,c\pm id$& $\zeta_{3}$&ST \\ \hline\end{tabular}

\vskip 0.5in
\begin{center}
 {\bf TABLE 2}\\
\vskip 0.1in
\begin{tabular}{|l|c|c|c|}
\hline
Experiment & Value of laminar scaling  exponent & CML exponent $\zeta$ &
Range o
f $\zeta$ \\  \hline
Convection in an annulus & $1.9 \pm 0.1$ & $\zeta_1$ &$1.9-2.2$ \\
\hline
Roll coating system &$ 0.63 \pm 0.02$ & $\zeta_3$ & $0.61-0.72$  \\
\hline
Convection (channel) & $1.6 \pm 0.2$ & $\zeta_G$ & $1.62-1.65 $  \\
\hline
Taylor Dean system & $1.67 \pm 0.14$ & $\zeta_G$ & $1.62-1.65$  \\
\hline
\end{tabular}
\vskip 0.5in
\end{center}

\newpage
\section {Figure Captions}
\begin{enumerate}

\item {\bf Fig 1(a)} shows the spatial intermittency in the lattice 
associated with tangent -period doubling bifurcation. {\it i} refers 
to position of site on the lattice.{\bf Fig 1(b)} shows two iterates of 
the lattice. As can be seen both regular and irregular sites behave 
periodically in time.The fixed point is $\xi={{\gamma\epsilon-\mu-2}\over{\mu-\gamma\epsilon}} $ 
\vskip 0.15in
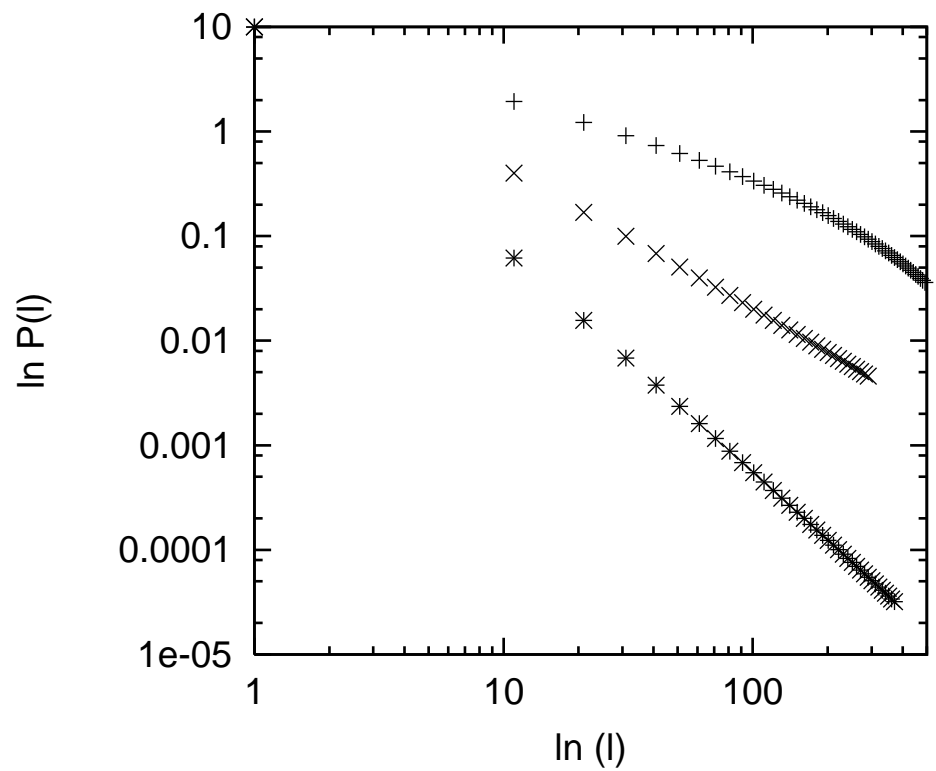
\item {\bf Fig 2} The scaling behaviour of the laminar sites is shown. $P(l)$ is 
the probability of obtaining a laminar region of length $l$.
The length of the laminar region is defined as number of adjacent sites 
which remain within a particular accuracy.
Three different exponents are found corresponding to three kinds of 
intermittency.The fixed point is 
$\xi={{\gamma\epsilon-\mu-2}\over{\mu-\gamma\epsilon}} $ .The behaviour observed was obtained for a lattice of size 50,000 iterated for 20000 iterates starting with $100$ random initial conditions. The accuracy to which the laminarity of the region was 
checked 
was $10^{-5} $. The asterisks  denote the behaviour
 with exponent $\zeta_{1} $, crosses $\zeta_{2}$ and pluses
denote $\zeta_{3}$. The ranges of the exponents are given in the caption of Table I.  $\zeta_3$ shows departures from this power-law over the third decade. The axes
are marked in the natural log-scale.

\vskip 0.15in
\item {\bf Fig 3}
The spatial second return map 
$x(i+2)$ vs $x(i)$ where
$i$ is the site index
is plotted for $\xi={{\gamma\epsilon-\mu-2}\over{\mu-\gamma\epsilon}}$.

\end{enumerate}

\section{Table Captions}
\begin{enumerate}
\item{\bf Table 1.1} We list the bifurcations from the synchronised
fixed point $x=0$. The type of bifurcation involved,the region in parameter space
 where the bifurcation conditions are satisfied, the manner in which the eigenvalue 
crosses unit circle and the power law exponent $P(l)\approx l^{\zeta_{i}}$
are listed.
The range of the exponent 
$\zeta_{1}$ is $[1.9-2.2]$, $\zeta_{2}$ lies in the range $[1.3-1.35]$ and
 $\zeta_{3} $ in the range $[0.61-0.72]$.
We also identify  the nature  of the intermittency, whether spatial or spatio-temporal(ST). 
In  {\bf Table 1.2} the same quantities are given for the other fixed point.

\item{\bf Table 2} This table lists the values of the spatial laminar exponent 
observed in experiments (values as quoted in Ref. \cite{Colovas}) and compares their values with the laminar exponents of 
our CML model. It is clear that the exponents $\zeta_1$ and $\zeta_3$ are directly observed in fluid experiments on convection in an annulus and in the roll coating system. As explained in the text, the exponent $\zeta_2$ serves as a lower bound on the ex
ponent observed in convection in a channel and in the Taylor Dean system. The
experimentally observed exponent actually coincides with the 
CML exponent $\zeta_G$ which occurs when the structure associated with the exponent $\zeta_2$ undergoes a further bifurcation. 
\end{enumerate}

\end{document}